# Striving for Realism, not for Determinism:
# Historical Misconceptions on Einstein and Bohm

Flavio Del Santo

*Institute for Quantum Optics and Quantum Information, Austrian Academy of Science, Vienna*
*Basic Research Community for Physics (BRCP)*

*Abstract* – In this paper I show that, while Einstein and Bohm both pursued a deterministic description of quantum mechanics, their philosophical concern was in fact primarily *realism* and not *determinism*. Their alleged firm adherence to determinism is based on a long-lasting misleading narrative and it should therefore be greatly scaled down.

**Introduction**

It has recently been critically pointed out that

> the most familiar fact about Einstein and quantum mechanics is that *he just didn't like it*. He refused to use the theory in its final form. And troubled by the fundamental indeterminism of quantum mechanics, he famously dismissed its worldview with the phrase "God does not play dice.[1]" (Stone, 2013, p. 2)

In the popularised (and sometimes vilified) narrative, this quoted phrase represents the very core of a misleading historiographical portrait of Einstein as being incapable of accepting the incredibly fortunate theory which he contributed to develop in the old days. In fact, "Einstein had turned, in the eyes of many working physicists, from revolutionary to reactionary, and his later views were considered curious at best" (Sauer, 2013).

In recent years, however, the tide turned in the historical role attributed to Einstein's contribution to quantum mechanics (QM). Along these lines, the present paper aims at clarifying certain misconceptions of Einstein's ideas on determinism, in relation to David Bohm who developed the first (and hitherto the only) deterministic interpretation of quantum mechanics (Bohm, 1952).

Concerning Einstein's relevance for QM, a historiographical breakthrough was achieved by Thomas Kuhn (1978), who firstly acknowledged Einstein to have been the real initiator of quantum physics (as opposed to Planck in 1900). In fact, in a most famous paper "On a Heuristic Point of View Concerning the Production and Transformation of Light" (Einstein, 1905),[2] Einstein firstly attributed physical reality to *light quanta* (*Lichtquanten*) and he himself evaluated this contribution – the only case in his outstanding career – as "most revolutionary" (see e.g. Jammer, 1992). In more recent years, this revaluation of Einstein's pivotal contributions to quantum physics is finally getting over the standard story. In this respect, Olivier Darrigol showed that Einstein was "the true discover of the quantum discontinuity" (Darrigol, 1992), whereas Helge Kragh noticed that besides Planck, "other physicists, and in particular Einstein, were crucially involved in the creation of quantum theory" (Kragh, 2000). But the most drastic historiographical novelty is a reappraisal of Einstein's late critiques of quantum theory, mostly due to Arthur Fine (1986) and Don Howard (1985, 1990), who firstly pointed out that Einstein actually did not write the famous EPR argument against the completeness of quantum theory (Einstein et al, 1935), confessing to Schrödinger that the paper "was written [...] by Podolsky, after many discussions. It did not come out in the end so well [...]; rather the main point was, so to speak, buried by erudition" (see Howard, 1985). More recently, Tilman Sauer (2013) and Stone (2013), noticed that only for Einstein's contributions to quantum physics "four Nobel Prizes would be about right".[3]

---

[1] This most famous quotation was written by Einstein in a letter to his colleague and friend Max Born in 1926 and it reads: "The theory says a lot, but does not really bring us any closer to the secret of the 'old one'. I, at any rate, am convinced that *He* is not playing at dice" (in Born, 1971, p. 91).

[2] This paper, one of the four great contributions of his *annus mirabilis*, is best known for Einstein's explanation of the photoelectric effect – for which he was awarded the Nobel Prize in 1921 – but it is actually the first conceptual breakthrough that paved the way to quantum physics.

**Determinism, realism and hidden variables**

As it is generally known, classical physics (i.e. Newton's and Maxwell's equations) gives deterministic predictions, and this invariably led many generations of physicists to undoubtedly believe that natural laws are *deterministic* relations. We call under the name of determinism

> the view that a sufficient knowledge of the laws of nature and appropriate boundary conditions will enable a superior intelligence to predict the future states of the physical world and to retrodict its past states with infinite precision. (Falkenburg and Weinert, 2009)

However, notice that a fixed cause-effect relation does not necessarily entail determinism (inasmuch as a natural law could be inherently probabilistic). Thus, while determinism implies causality, the inverse *non sequitur* (a concrete example will be provided below with Popper's propensity interpretation of probability).

Incidentally, it should be noticed that textbooks tend to consider the advent of statistical mechanics as no threat to determinism in so far as the stochastic character is attributed to a lack of computational power, or of knowledge (epistemological indeterminism), and is not regarded as a fundamental characteristic of nature (ontological indeterminism). However, it ought to be remarked that Ludwig Boltzmann – who perhaps more than anyone else contributed to develop statistical mechanics – has been the first modern physicist to challenge the idea of determinism as early as 1895 (see Jammer, 1973). Also Franz Exner, Boltzmann's successor in Vienna, took an even more radical position, proposing, in 1908, that also at the macroscopic level determinism is only apparent:

> in the region of the small, in time and space, the physical laws are probably invalid; the stone falls to earth and we know exactly the law by which it moves. Whether this law holds, however, for each arbitrarily small fraction of the motion [...] that is more than doubtful (see Hanle, 1979)

And Schrödinger, who served as Exner's research assistant in Vienna, "most firmly embraced [Exner's] fundamental indeterminacy before quantum mechanics." (Hanle, 1979, p. 257).

But it was only with the onset of quantum physics that indeterminism became widely discussed. In 1900, in "an act of desperation", Max Planck used the stratagem of dividing into discrete packages the energy that light can exchange in a perfectly absorbing body (*black body*). As already recalled, it was Einstein who firstly used the term *quanta* and, basically alone for a decade, he strove for a double corpuscular-undulatory ontology of light. Moreover, it was again Einstein who firstly took quanta out of the domain of solely optical phenomena into matter (Einstein, 1907) and, most remarkably, Einstein himself introduced genuine *randomness* in emission processes of single quanta of light (Einstein, 1916; see also Stone, 2013, pp. 186 ff.).

---

[3] Worth mentioning is that the excellent and most influential book by Max Jammer, *The Philosophy of Quantum Mechanics* (Jammer, 1974), helped corroborating the portrait of a reactionary Einstein who lost the debate with Bohr. However, in a later work (Jammer, 1992), the author fully recognised the central importance of Einstein: "It is generally given for granted that the theories of relativity are mainly the result of Einstein's work. However, it is not likewise well-known that the development of quantum physics is greatly indebted to Einstein".

But then, what was Einstein's uneasiness with quantum physics? To understand this, we shall take a step further into the theory.

Since 1926, the main element of QM is the so-called wave-function (or quantum state), $\Psi$, a mathematical object that encompasses the "physics" of a system. The time evolution of this wave-function is described by the Schrödinger equation. However, quantum formalism allows only to compute the probabilities for different outcomes to occur in an experiment, and there is no way to predict which of them will turn up. Namely, QM gives, in general, only indeterministic predictions.

But this is not the whole story; QM also challenges the concept of realism. While quantum objects show both an undulatory and a corpuscular nature, a specific choice of the experimental setting reveals either one or the other of these natures but never both.[4] It thus seems that the choice of an *observer* actively influences the system under investigation, or as sometimes is put, the observer "realises" (one or another) reality upon measurement. In the words of Werner Heisenberg, one of the fathers of orthodox interpretation, "the conception of objective reality […] has thus evaporated" (Heisenberg, 1958).

The current probabilistic interpretation of the Schrödinger equation was put forward by Max Born in 1926 – for which he won the Nobel Prize – and, again, this was admittedly inspired by Einstein (see Pais, 1982, p. 1196). In his pivotal work, Born wrote:

> this raises the whole problem of determinism. From the standpoint of our quantum mechanics, there is no quantity that could establish the effect of a collision causally in the individual cases; however, up to now, we have no clue regarding the fact that there are internal properties of the atom […]. I myself tend to abandon determinism in the atomic world." (Born, 1926).

These "internal properties" that could in principle deterministically complete QM are referred to as *hidden variables*. It is not well known that, stimulated by the newly introduced Born's interpretation,

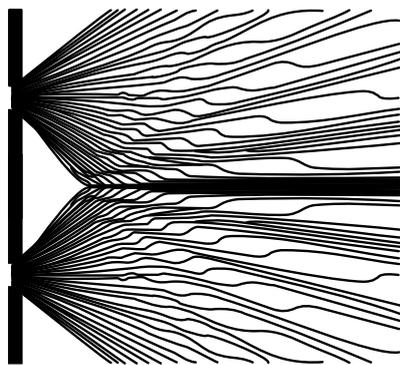

*Figure 1:* Trajectories of particles in the double slit experiment in Bohm's hidden variable interpretation.

Einstein himself devised a hidden variable model as early as 1927. This model allowed to determine, given the knowledge of certain *hidden* parameters, the actual trajectories followed by particles.[5] In this paper, however, Einstein is reluctant to explicitly give a realistic interpretation to $\Psi$, because, as he noted before, a function living in a "many-dimensional coordinate space does not smell like something real" (see Howard, 1990, p. 83). And he strengthened his criticism stating that "there are objections in principle against this multi-dimensional representation" (see Belousek, 1996, p. 455). Einstein himself prevented his paper from appearing in print, and Belousek showed how in the following years Einstein became increasingly critical towards any further attempt of synthesising the quantum and wave conceptions. Even when the American physicist David Bohm developed his celebrated deterministic *hidden variables* model (Bohm, 1952) which reproduces all the predictions of QM,[6] Einstein maintained this untenable. So, despite Bohm's model virtually achieved the goal of Einstein's 1927 paper, the latter wrote to Born:

> have you noticed that Bohm believes (as de Broglie did, by the way, 25 years ago) that he is able to interpret the quantum theory in deterministic terms? That way seems too cheap to me". (Einstein to Born, 12.05.1952. Born, 1971, p. 192).

---

[4] This concept is at the basis of Niel Bohr's complementarity interpretation, which, in turn, provides the philosophical foundations of the mainstream Copenhagen interpretation of QM. As already recalled Einstein strove for a double nature of light, but as a simultaneous and not exclusive feature, and in this he inspired Luis de Broglie when, in 1923, he advanced the hypothesis of a double behaviour, undulatory and corpuscular, of matter (i.e. electrons, neutrons, etc.) as well. However, Howard stressed that "it is wrong to credit the idea of material particles possessing simultaneously a wave-like aspect wholly to de Broglie, since Einstein was well-known even at that time to have toyed with such ideas since at least 1921" (Howard, 1990). These ideas were revived in post-war period by the physicists David Bohm (see main text), Jean-Pierre Vigier, Franco Selleri (see Baracca *et al.*, 2017) and others.

[5] Einstein presented this result at the Prussian Academy of Science in Berlin in May 1927. However, he never published his results, which survived in the form of a manuscript (see Belousek, 1996). The hidden variables of this model, which is restricted to time-independent Hamiltonian evolutions, are $n$ real, distinct scalars that determine $n$ mutually orthogonal directions in configuration space, where $n$ is the number of degrees of freedom.

What we learn from Einstein's involvement in the hidden variable programme, is that his main concern was definitely not determinism. His own and de Broglie's early incomplete attempts, and Bohm's consistent interpretation all achieved to retrieve determinism, however they were not enough for Einstein. The fact that they all relied on a wave-function living in a configuration space, made them despicable to Einstein, in so far as they clearly did "not smell like something real". Sacrificing a tenable form of realism was a too high price to pay for Einstein, even if determinism was so restored. Einstein was surely in favour of determinism, but this was definitely not a fundamental requirement to him.

In this respect, Karl Popper put forth the view that physics (both classical and quantum) is based on realism, but it is fundamentally indeterministic. In 1950, Popper presented in Princeton his ideas on indeterminism and Einstein and Bohr attended. In discussions, Popper tried "to persuade [Einstein] to give up his determinism" (Popper, 1976, p. 148). Yet, later he stated:

> the attribution to Einstein of the formula 'God does not play with dice' is a mistake. Admittedly, he was a strict determinist when I first visited him in 1950 […]. But he gave this up […]. Einstein was, in his last years, a realist, not a determinist." (See Del Santo, 2018)

Such a claim is supported by the authoritative judgement of Wolfgang Pauli, who wrote to Born:

> Einstein does not consider the concept of 'determinism' to be as fundamental as it is frequently held to be (as he told me emphatically many times) […] Einstein's point of departure is 'realistic' rather than 'deterministic', which means that his philosophical prejudice is a different one. (31.03.1954 letter from W. Pauli to M. Born, reproduced in Born, 1971, p. 221).

As an alternative to determinism, Popper developed, in late 1950s, the idea that there exist fields of *propensities*, i.e. real waves of probability – characteristic of the whole experimental arrangement – which determine the distribution of outcomes of experiments (see Popper, 1982, p. 64 ff.). To this extent, "the term *propensity* refers to a degree of *causality* that is weaker than determinism" (Ballentine, 2016), but that preserves realism. Unfortunately, Einstein died two years before the first formulation of propensities, and we will never know his opinion on this interpretation.

On the other hand, given Popper's aversion towards determinism, he also did not champion Bohm's interpretation. In the book that collects his mature views on foundations of QM, Popper indeed stated:

> in spite of Bohm's realist and objectivist programme, his theory is unsatisfactory […]. It is not only bound, like all other deterministic theories, to interpret probabilities subjectively, but it even retains Heisenberg's 'interference of the subject with the object' […]. (Popper 1982, p. 174).

---

[6] In Bohm's model, quantum particles follow deterministic trajectories (see Fig. 1), but the probabilistic behaviour of quantum mechanics is given by the impossibility of knowing the initial position of particles (which remains a hidden variable).

In an unpublished letter, however, Bohm firmly replies to Popper that, like Einstein, his main concern was realism and that determinism was used merely instrumentally:

> I certainly think that a realistic interpretation of physics is essential. […]. However, I feel that you have not properly understood my own point of view, which is much less different from yours than is implied in your book. Firstly I am not wedded to determinism. It is true that I first used a determinist version of […] quantum theory. But later, […] a paper was written,[7] in which we assumed that the movement of the particle was a stochastic process. Clearly that is not determinism. […] The key question at issue is therefore not that of determinism vs. indeterminism. I personally do not feel addicted to determinism, but I am ready to consider deterministic proposals, […] if they offer some useful insights. (D. Bohm to K. Popper 13.07.1984. PA, 278/2).

In conclusion, contrarily to the standard story (especially widespread among physicists), neither Bohm nor Einstein were staunchly committed to determinism and they would have accepted fundamental indeterminism in exchange for realism.

**Acknowledgements**

I wish to express my gratitude to Prof. Jürgen Renn for providing me the opportunity of a visiting fellowship at the Max Plank Institute for the History of Science in Berlin, where this paper was written. I am also thankful to Prof. Angelo Baracca and Mr. Roberto Amabile for useful comments.

---

[7] The mentioned paper is (Bohm and Vigier, 1954).